\documentclass{aa}

\authorrunning{Caunt \& Korpi}
\titlerunning{3D MHD model of astrophysical flows}

\usepackage{graphics}

\begin{document}

\newcommand\vect[1]{{\mbox{\boldmath $#1$}}}
\newcommand{\cold}{{_{\rm c}}}
\newcommand{\w}{{_{\rm w}}}
\newcommand{\h}{{_{\rm h}}}

%
%
\newcommand{\cm}{\,{\rm cm}}
\newcommand{\mm}{\,{\rm mm}}
\newcommand{\cmcube}{\,{\rm cm^{-3}}}
\newcommand{\dyn}{\,{\rm dyn}}
\newcommand{\erg}{\,{\rm erg}}
\newcommand{\Jy}{\,{\rm Jy}}
\newcommand{\Jyb}{\,{\rm Jy/beam}}
\newcommand{\kms}{\,{\rm km\,s^{-1}}}
\newcommand{\mJy}{\,{\rm mJy}}
\newcommand{\mJyb}{\,{\rm mJy/beam}}
\newcommand{\K}{\,{\rm K}}
\newcommand{\kpc}{\,{\rm kpc}}
\newcommand{\Mpc}{\,{\rm Mpc}}
\newcommand{\mG}{\,{\rm mG}}
\newcommand{\mkG}{\,\mu{\rm G}}
\newcommand{\MHz}{\, {\rm MHz}}
\newcommand{\Msol}{\,{\rm M_\sun}}
\newcommand{\p}{\,{\rm pc}}
\newcommand{\radm}{\,{\rm rad\,m^{-2}}}
\newcommand{\s}{\,{\rm s}}
\newcommand{\yr}{\,{\rm yr}}

\title{A 3D MHD model of astrophysical flows: algorithms, tests 
and parallelisation}

\author{S.~E. Caunt and M.~J.\ Korpi}
\institute{Astronomy Division, Department of Physical Sciences,
  P.~O.\ Box 3000, FIN-90014 University of Oulu, Finland}

\offprints{M.~J.\ Korpi, \email{Maarit.Korpi@oulu.fi}}

\abstract{In this paper we describe a numerical method designed for
modelling different kinds of astrophysical flows in three
dimensions. Our method is a standard explicit finite difference method
employing the local shearing-box technique.
To model the features of astrophysical systems, which are usually
compressible, magnetised and turbulent, it is desirable to have high
spatial resolution and large domain size to model as many features as
possible, on various scales, within a particular system. In addition, the
time-scales involved are usually wide-ranging also requiring
significant amounts of CPU time.
These two limits (resolution and time-scales) enforce huge limits on
computational capabilities. The model we have developed therefore uses
parallel algorithms to increase the performance of standard serial
methods. The aim of this paper is to report the numerical methods we
use and the techniques invoked for parallelising the code. The
justification of these methods is given by the extensive tests
presented herein. 
}

\maketitle


\section{Introduction}

Magnetic fields are present everywhere in the universe, for example in
planets, stars, accretion disks around compact objects, galaxies of
various kinds and even in the intergalactic medium. Most of these
systems are characterised by large kinetic and magnetic Reynolds
numbers, indicating that they are highly turbulent, and also that
magnetic fields are dynamically significant. In many cases the
observed magnetic field has, in addition to a random small-scale
component, coherent magnetic field structures on large scales. For
example the Sun has a mean magnetic field of dipolar structure,
whereas numerous spiral galaxies posses a mean field of spiral shape
often following the optical spiral arms. One of the fundamental
questions in astrophysics is how the order seen in the large scale
magnetic field structures can arise in the turbulent media they are
embedded within.

The most plausible mechanism suggested to explain this phenomenon is
the hydromagnetic dynamo (Parker \cite{Parker55}; Steenbeck et
al. \cite{SKR66}), according to which large-scale magnetic fields can
be generated and maintained by the combination of turbulence and
large-scale shearing motions. Based on this theory, a vast number of
mean-field dynamo models, which solve for the large scale magnetic
field with turbulence remaining a parameterised quantity, have been
developed for practically all astrophysical objects. Since the form
and magnitude of the turbulent quantities are relatively unknown, this
parameterisation is usually kept as simple as possible. The
information lacking from these models can be obtained by studying the
non-linear evolution of a magnetised turbulent flow in a fully 3D
numerical simulation. These kind of simulations (e.g. Balsara \&
Pouquet \cite{Balsara99}; Brandenburg \cite{Aksu2000}) have revealed
that in the presence of helical turbulence, magnetic field energy can
be transferred from the smallest scales to the larger ones, known as
inverse cascade (Frisch et al. \cite{Frisch75}; Pouquet et
al. \cite{inverse76}). These simulations, however, have not yet
developed to the stage where a realistic physical setup for a
particular object could be studied.

On the other hand, when the magnetic field becomes strong enough it
can influence the fluid motions through the Lorentz force suppressing
turbulence and thereby quenching the generation of magnetic field,
known as $\alpha$-quenching in the mean field dynamo theory.  Recent
numerical simulations (Cattaneo \& Hughes \cite{Hughes96}) have
indicated that the dynamo $\alpha$ is dramatically quenched implying
that dynamo action cannot occur in high Reynolds number
flows. However, this result is still debatable and requires
investigation under more realistic physical setups (see for example
Brandenburg 2000).

Modelling these kinds of systems provides a wide range of numerical
challenges. One challenge that can never be overcome satisfactorily is
the need for the highest possible spatial resolution to model
turbulence. In some cases even the turbulent forcing occurs at very
small scales, for example in galaxies where turbulence is mainly
driven by dying stars exploding as supernovae (hereafter SNe). On the
other extreme, one would also like to include the largest possible
scales to study not only the generation of large-scale magnetic field,
but also large-scale vertical structures such as chimneys or fountains
observed in galaxies (e.g. Koo et el. \cite{Koo92}; Normandeau et
al. \cite{Norma96}) and azimuthal features such as field reversals in
accretion disks (e.g. Brandenburg et al. \cite{BNST95}). Another
important feature of astrophysical flows which has to be taken into
account is their compressible nature. With the violent physical
processes active in these flows, such as SNe in galaxies and rapidly
growing instabilities like the Balbus-Hawley instability in accretion
disks (Balbus \& Hawley \cite{BH91}), shocks are commonly
formed. Since the physical viscosity in the flow is negligible and the
physical quantities become discontinuous, additional numerical
techniques are required to resolve them (von Neumann \& Richtmyer
\cite{vonNeumannR50} hereafter vNR). Once the flow has become
turbulent, the time-scales involved in these turbulent motions is
usually much shorter than the orbital period or decay time-scale of the
magnetic field. To study the long-term evolution of the system a huge
number of time-steps may be required. This all implies the need for
efficient algorithms along with high resolution and large domain size.

A number of numerical models of this type have been developed being
either specifically designed for a particular object (for solar corona
e.g. Galsgaard \& Nordlund \cite{GalsgaardN96}, for stellar convection
e.g. Stein \& Nordlund \cite{SteinN98}, for accretion disks
e.g. Hawley et al. \cite{HawleyGB95}; Brandenburg et
al. \cite{BNST95}, for the interstellar matter e.g. Rosen \& Bregman
\cite {RosenB95}; V\`azquez-Semadeni et al. \cite{VazquezP95}; Mac Low
\cite{MacLow99}; Korpi et al. \cite{Korpi99}) or being more general
(Stone \& Norman \cite{StoneN92}, \cite{StoneN92b}), and more recently
global models are starting to appear (e.g. for the accretion disks
Hawley \& Krolik \cite{HawleyK00}). Our method is based on the
standard local Cartesian shearing-box simulation (e.g. Wisdom \&
Tremaine \cite{WisdomT88}) and uses explicit finite differences on an
Eulerian grid, discretising physical quantities onto a uniform mesh,
ideal for data parallelisation. The methods are in principle very
simple and therefore easy to implement and allow for rapid
development. Shock viscosities (vNR) and further diffusive techniques
(Nordlund \& Galsgaard \cite{NordlundG97} hereafter NG; Stein \& Nordlund
\cite{SteinN98}) require additional effort but are however necessary
to stabilise the numerics.

This paper is structured as follows.  Sect. 2 describes the
essential physics behind our model. Sect. 3 provides details of the
numerical methods we use to model the fluid including the artificial
viscosity employed to resolve shocks and reduce unphysically generated
waves and the treatment of the boundaries of the box. The
parallelisation of the code is discussed in Sect. 4 and finally the
test suite used to verify the accuracy and acceptability of the code
is covered in Sect. 5. Finally in Sect. 6 we summarise.



\section{The generalised model}

\subsection{Introduction}

In this section we discuss the partial differential equations (PDEs)
solved for in all astrophysical systems under consideration. These
equations describe the flow of a magnetically conducting fluid within
a differentially rotating body. Other, more specific, models include
extra terms to model effects such as heating by supernovae (hereafter
SNe) or stellar winds (hereafter SWs), and suitable cooling
functions for various systems.

\subsection{The non-ideal MHD equations}
\label{s:eqns}

We solve the standard non-ideal MHD equations in three dimensions
using the standard shearing-box technique. The equations are solved in
a computational domain representing a small volume within a
differentially rotating cosmic object. The coordinate system is
reduced to Cartesian with $x$ representing radial, $y$ azimuthal and
$z$ vertical direction, the dimensions of the box being $L_x \times
L_y \times L_z$. The centre of the box is located at distance
$R$ from the centre of rotation, which is much larger than any
dimension of the domain. This reference point is moving on a circular
orbit with angular velocity $\Omega_0$ around the centre. As the fluid
is rotating differentially, the angular velocity is changing as
function of distance from the centre of rotation. In the local frame
of reference the equations of motion can be linearised relative to the
reference point (e.g. Spitzer \& Schwarzschild \cite{SS53}; Julian \&
Toomre \cite{JT66}) yielding a solution which can be interpreted to
have two contributions: the circular motion given by $u_0=-2 A x
\hat{y}$ and the epicyclic motion which yields a term $-2 A u_x
\hat{y}$, where $A=-1/2 R \left( \partial \Omega / \partial x
\right)_{R}$ is equivalent for the Oort constant.  For disk systems,
for example, for which the rotation law is of the form $\Omega_0
\propto R^{-q}$, the Oort constant $A$ can be written as $1/2 q
\Omega_0$, yielding the general form ${\bf u}_0=-q \Omega_0 x {\hat
y}$ for the shear flow and $-q \Omega_0 u_x \hat{y}$ for the epicyclic
motion. Our velocity field then consists of two parts, the shear flow
${\bf u}_0$ discussed above, and deviations ${\bf u}$ from it, the
total velocity field being ${\bf U}= {\bf u}_0 + {\bf u}$. In the
following we solve for ${\bf u}$.

We choose to solve for the magnetic vector potential, ${\bf A}$, for
which $\nabla \cdot {\bf B}=0$ is a natural consequence. We solve for
the internal energy $e$, which is related to temperature by $e=c_v T$
under the assumption of the perfect gas law $p = \rho e (\gamma-1)$
with $\gamma=c_p/c_v=5/3$.  Finally we solve for the logarithm of
density $\ln{\rho}$, which is numerically convenient, since the
density range can be several orders of magnitude in many
models. However, this is not a conservative form of the continuity
equation, but the extensive tests have shown this not to be a major
disadvantage. Other factors, such as open boundaries and numerical
diffusion, would destroy the conservative nature of the continuity
equation even if $\rho$ was solved for. The basic equations we solve
are
\begin{eqnarray}
   \frac{\partial {\bf A}}{\partial t} &=& 
    {\bf U} \times {\bf B} - \eta \mu_0 {\bf J}, \label{e:ind} \\ 
   \frac{\partial {\bf u}}{\partial t} 
   &=&
     -({\bf U} \cdot \nabla) {\bf u}
     - \frac{1}{\rho} \nabla p
     - 2 {\bf \Omega} \times {\bf U}
     - q \Omega_0 u_x \hat{y} \\ \nonumber
     &&\hspace{2cm}+ {\bf g}
     + \frac{1}{\rho} {\bf J} \times {\bf B} 
     + \frac{1}{\rho} \nabla \cdot {\bf \tau}, \label{e:mom} \\
   \frac{\partial e}{\partial t}
   &=&
     - \left( {\bf U} \cdot \nabla \right) e
     - \frac{p}{\rho} \left( \nabla \cdot {\bf U} \right)
     + \frac{1}{\rho} \nabla \left( \chi \rho \nabla e \right) \\ \nonumber
     &&\hspace{2cm}+ Q_{\rm visc}
     + Q_{\rm Joule}, \label{e:ene} \\
   \frac{\partial \ln\rho}{\partial t}
   &=&
     -\left( {\bf U} \cdot \nabla \right) \ln\rho 
     - \nabla \cdot {\bf U}, \label{e:con}
\end{eqnarray}

\noindent 
Here ${\bf J}=\mu_0^{-1}\nabla\times{\bf B}$ is the current density,
$\mu_0$ the permeability of free space, $\tau$ the stress tensor,
$Q_{\rm visc}$ the viscous dissipation, and $Q_{\rm Joule}$ the Joule
dissipation. The diffusion terms involving the stress tensor, $\tau$,
magnetic diffusivity, $\eta$, and thermal diffusivity, $\chi$, are
included in the equations to emphasise that diffusion is incorporated
into the model, however these are treated as purely numerical
operations. It should also be noted that a diffusion term is
incorporated in the continuity equation to stabilise the model. A
detailed discussion of the diffusive terms is covered in Sect.
\ref{s:artvis}.

The term ${\bf g}$ in the momentum equation describes the external
gravitational potential. For accretion disks, for example, we estimate
the gravity by linearising the equation of motion, which yields gravity
in the vertical direction $g_z=- \Omega_0^2 \hat{z}$. We neglect
self-gravity for the time being. 

Additional specific terms to Eqs. (\ref{e:mom}) to (\ref{e:con}) are
included for different simulations, for example, inclusion of
turbulent forcing mechanisms and appropriate cooling
functions. However, the equations above are common throughout the
models and lay the foundations for all future calculations.

\subsection{Boundary conditions}
\label{s:bound}

In the azimuthal, $y$, direction we adopt periodic boundary conditions
since this lies in the direction of the shearing flow. In the radial
direction, the differential rotation and therefore shearing boundaries
need to be accounted for. For the linear shear we therefore adopt
\begin{equation}
f(L_x,y,z)=f(0,y+ q \Omega_0 L_x t,z),
\end{equation}
where $f$ represents any of the eight variables.  Since the effect of
shearing, $q \Omega_0 L_x t$ typically yields a position that does not
lie directly on a grid-point, further interpolation is required at the
boundaries to account for this. The exact implementation of this is
discussed in Sect. \ref{s:numerics}. 

In the vertical direction we have two schemes available. Since these
boundaries are the hardest to model since they are not `true'
boundaries in a physical system, we must chose conditions which best
suit the particular physical situation to be modeled. We always,
however, assume stress-free, electrically insulating boundary
conditions, such that
\begin{eqnarray}
\frac{\partial A_x}{\partial z}=\frac{\partial A_y}{\partial
z}=A_z &=& 0,\\
\frac{\partial u_x}{\partial z}=\frac{\partial u_y}{\partial z}&=&0,\\
\frac{\partial e}{\partial z}&=&0.
\end{eqnarray}
We then employ either `open' or `closed' boundaries by setting 
\begin{equation}
\frac{\partial u_z}{\partial z}=0,
\end{equation}
for open boundaries and 
\begin{equation}
u_z = 0,
\end{equation}
for closed. The boundary condition for density comes from hydrostatic
equilibrium at the surfaces yielding
\begin{equation}
\frac{\partial \ln\rho}{\partial z} = \frac{g}{\left( \gamma-1 \right)
e}.\label{e:rhobc}
\end{equation}
The numerical implementation of these boundary conditions is
discussed in the following section.



\section{The numerical methods}
\label{s:numerics}

\subsection{Introduction}

Our code is based on explicit finite difference calculations
using an array of data of size $n_x \times n_y \times n_z$ uniformly
spaced gridpoints. We numerically solve for the eight primary
variables $\rm{ln}\rho$, $e$ and components of ${\bf u}$ and ${\bf
A}$ which represent the logarithm of density, energy per unit mass,
velocity and the magnetic vector potential. 

We use the logarithm of density for a number of reasons. It ensures
that we never obtain negative densities, it allows us to cope with
physical situations that require a large number of pressure scale
heights and finally the functional form of $\ln\rho$ is much smoother
than that of $\rho$, hence numerical derivatives are more accurately
calculated.

The discretisation of the partial derivatives in $x$, $y$ and $z$ are
done using centred, 6th order accurate, explicit finite differences
for both first and second derivatives. The exact form of these is
included in Appendix \ref{a:findif}. These operators are highly
non-dissipative with well defined waves retaining their original form
over long periods of time. 

Time-stepping is performed by a third order accurate
Adams-Bashforth-Moulton predictor-corrector method, which is described
in Appendix \ref{a:time-steps}. The accuracy of this scheme has been
compared to other methods of advancing PDEs as discussed in Sect.
\ref{s:tscomp}.

\subsection{Numerical diffusion}
\label{s:artvis}

The methods described above for solving the system of PDEs are
inadequate alone to cope with strong discontinuities in the flow, such
as shock waves, and are susceptible to low-level numerical noise. We
therefore employ artificial viscosities to diffuse the discontinuities
to be resolved by the finite computational grid and add stability to
the numerical methods.

The methods we use generate viscosities that are localised at
discontinuities or in regions of unresolved waves. This means that we
are able to apply the minimum amounts of viscosity to those areas in
which we would like the flow to remain unchanged. We use two
techniques to account for these different numerical problems: a shock
viscosity (vNR) and hyperdiffusion (NG). 

We use artificial counterparts to the physical quantities of $\nu$,
$\eta$ and $\chi$ being the kinematic viscosity, magnetic diffusivity
and thermal diffusivity, respectively. The numerical equivalent of
$\nu$ is incorporated into the stress tensor and viscous heating as
discussed in Sect. 3.2.3, and $\eta$ and $\chi$ are the numerical
equivalents of quantities in the Eqs. (\ref{e:ind}) and (\ref{e:ene}).

\subsubsection{Shock viscosity}

The shock viscosity is only applied to regions that are undergoing
compression, i.e. in regions which are characterised by
$\nabla\cdot{\bf u} < 0$. The numerical equivalent of the kinematic
viscosity $\nu$ therefore takes the form of

\begin{equation}
\nu_i^{\rm shk}=\left\{
\begin{array}{ll}
c_{\rm shk}\Delta x_i^2|\nabla\cdot{\bf u}| & \nabla\cdot{\bf u} < 0 \\
0 & \nabla\cdot{\bf u} \ge 0
\end{array}
\right.,
\label{e:shockvis}
\end{equation}

\noindent where the effect of $c_{\rm shk}$ is to produce greater
damping of the shock resulting in spreading the shock over more
gridpoints and is typically of the order of unity. For the
thermal diffusivity, $\chi_i^{\rm shk}=\nu_i^{\rm shk}/Pr$.

A similar shock viscosity is required for the magnetic resistivity,
however we wish to ensure that diffusion only occurs from the
components of velocity perpendicular to the field lines, ${\bf
u_\perp}$, given by

\begin{equation}
{\bf u_\perp} = {\bf u}-\frac{({\bf u}\cdot{\bf B}){\bf B}}{|{\bf B}|^2}
\end{equation}

The form of the magnetic shock resistivity is then given in an identical
manner to the shock viscosity:

\begin{equation}
\eta^{\rm shk}_i=\left\{
\begin{array}{ll}
\frac{c_{\rm shk}\Delta_i^2}{P_M}|\nabla\cdot{\bf u_\perp}| & \nabla\cdot{\bf u_\perp} < 0 \\
0 & \nabla\cdot{\bf u_\perp} \ge 0
\end{array}
\right.,
\label{e:etashk}
\end{equation}

\noindent where $P_M$ is the magnetic Prandtl number. Hence for flows
along field lines, no magnetic diffusion occurs, while for field lines
that are strongly compressed into a small region by the flow this term
becomes large.

\subsubsection{Hyperdiffusion}

Hyperdiffusion is incorporated to add numerical stability to the
code. Small scale oscillations (around Nyquist frequency) need to
be damped and the hyperdiffusive methods described by NG provide an
efficient method while leaving resolved features practically undamped.

This is strongest for rapid (grid-scale) oscillations. In an
implementation termed `positive definite quenching' by NG the
hyperdiffusion always has physically meaningful values such that the
dissipation of energy is positive definite and always acts to
stabilise the flow. For a detailed description of the hyperdiffusive
techniques, we refer the reader to their article and Nordlund \& Stein
(\cite{NordlundS90}). Our implementation of the techniques is
discussed below.

Written in terms of viscosity, hyperdiffusion of a variable,
$f$, can be expressed as
\begin{equation}
\nu_i^{\rm hyp}(f) = c_{\rm hyp}\Delta x_i v q_i(\partial_if),
\label{e:hypervis}
\end{equation}

\noindent where $q_i(\partial_if)$ represents the hyperdiffusive
operator defined in the above references to be $q_i(f)=|\Delta^2
f|/|f|$ and

\begin{equation}
v=|{\bf u}|+c_s+v_A+|{\bf u_0}|
\end{equation}

\noindent taking into account the fluid velocity ${\bf u}$, the sound
speed $c_s=(\gamma P/ \rho)^{1/2}$, the Alfv\'en velocity $v_A = (|{\bf
B}|^2/\rho)^{1/2}$ and the underlying shearing flow ${\bf u_0}$ which,
as noted by Nordlund \& Stein (\cite{NordlundS90}) and Stein \&
Nordlund (\cite{SteinN98}), stabilises weak waves (sound waves and
fast mode waves) and prevents ringing at sharp changes in advected
quantities. 

Here $\nu$ represents a general viscosity term, and one can
equally substitute $\chi$ or $\eta$ for the energy and induction
equations, respectively, and mass diffusion in the continuity equation.

\subsubsection{Implementation of diffusive terms}

Eqs. (\ref{e:ind}) to (\ref{e:con}) all have additional diffusive
terms in order to stabilise the code. For the momentum equation, this
can be performed by replacing the stress tensor by a diffusive
operator (retaining the essential form of the stress tensor but using
numerical equivalents for viscosity). The magnetic diffusion
similarly is replaced by a numerical equivalent as does the thermal
diffusion. Mass diffusion however has no physical counterpart and is
included purely for stability.

The diffusive terms are calculated on a staggered mesh which provides
a more accurate method of determining grid-scale structures. This is
used in conjunction with second-order operators for determining highly
localised structures. This combination
allows high wavenumber noise to be detected more easily and
discontinuities to be dealt with more efficiently. 

The diffusion of the scalar quantities $e$ and $\ln{\rho}$,
represented by $f$ below, in the $i$th-direction can be written as
\begin{equation}
\frac{\partial{\it f}}{\partial {\it t}}=\dots
+\frac{1}{\rho}\partial^+_i(\nu^-_i(f)\rho^-\partial^-_i(f))
\label{e:scaldif}
\end{equation}
\noindent where the $+$ and $-$ signs indicate the direction in which
a particular operation is performed relative to a particular
grid-point, the final result being exactly on the grid-point and
$\nu_i(f)$ is defined simply to be the sum of the shock and
hypercomponents given by Eqs. (\ref{e:shockvis}) and
(\ref{e:hypervis}). For the energy equation, $\nu_i(f)$ can be
considered to be equivalent to $\chi$, the thermal diffusion
coefficient, and hence Eq. (\ref{e:scaldif}) can be regarded as an exact
numerical equivalent to the thermal diffusion approximation
\begin{equation}
\frac{\partial e}{\partial t} = \dots \frac{1}{\rho}\nabla
(\chi\rho\nabla e),
\end{equation}

\noindent and indeed a Prandtl number, $Pr$, is used to distinguish
this fact by assigning $\chi_i(f) = \nu_i(f)/Pr$.

The diffusion of the vector quantities is a more complex operation. In
the case of velocity, the diffusion is implemented in a way that
closely resembles the stress-tensor form of molecular viscosity
following the implementation illustrated by NG. In the momentum
equation we add a term of the form:
\begin{equation}
\label{e:difu}
   \frac{\partial {\bf u}}{\partial t} = \dots +
        \frac{1}{\rho} \frac{\partial}{\partial x_j}\tau_{ij}, 
\end{equation}

\noindent where $\tau_{ij}$ is the symmetrised stress tensor
\begin{equation}
\tau_{ij}=\frac{1}{2}(\epsilon_{ij}+\epsilon_{ji}), 
\end{equation}

\noindent and
\begin{equation}
\epsilon_{ij}=\rho \nu_j(u_i) \frac{\partial u_i}{\partial x_j},
\end{equation}

\noindent where $\nu_j(u_i)=\nu_j^{\rm shk}+\nu_j^{\rm hyp}(u_i)$. The
viscous dissipation feeds directly back into the energy equation by
defining
\begin{equation}
Q_{\rm diss} = \sum_{ij}\tau_{ij}\frac{\partial u_i}{\partial x_j}
\end{equation}

\noindent The role of positive definite quenching is noted here that,
through the definition of Eq. (\ref{e:hypervis}), this term remains
physically meaningful.

The magnetic diffusion is defined as
\begin{equation}
\frac{\partial {\bf A}}{\partial t} = \dots - {\bf E},
\end{equation}

\noindent where ${\bf E}=\eta \mu_0 {\bf J}$ and $\eta$ is a function
of ${\bf J}$ with direction dependency also and can be
expressed as

\begin{eqnarray}
E_x/\mu_0&=&   \eta_y^{\rm hyp}(B_z)\partial_y B_z
               +\eta_z^{\rm hyp}(B_y)\partial_z B_y \nonumber \\
&&\hspace{5mm}+(\eta_y^{\rm shk}+\eta_z^{\rm shk})J_x,  \nonumber \\
E_y/\mu_0&=&   \eta_z^{\rm hyp}(B_x)\partial_z B_x
               +\eta_x^{\rm hyp}(B_z)\partial_x B_z \nonumber \\
&&\hspace{5mm}+(\eta_z^{\rm shk}+\eta_x^{\rm shk})J_y, \nonumber \\
E_z/\mu_0&=&   \eta_x^{\rm hyp}(B_y)\partial_x B_y
               +\eta_y^{\rm hyp}(B_x)\partial_y B_x \nonumber \\
&&\hspace{5mm}+(\eta_x^{\rm shk}+\eta_y^{\rm shk})J_z.
\end{eqnarray}

\noindent Diffusion is taken in directions perpendicular to a
particular magnetic field component which is necessary to diffuse
those directions which contribute to the current. $\eta_i^{\rm shk}$
is taken from Eq. (\ref{e:etashk}) whereas $\eta_i^{\rm hyp}(f)$
follows that of Eq. (\ref{e:hypervis}) but is divided by the
magnetic Prandtl number, $P_M$ such that $\eta_i^{\rm hyp}(f) =
\nu_i^{\rm hyp}(f)/P_M$.

Finally, we have the additional term in the energy equation to account
for the losses in magnetic energy
\begin{equation}
Q_{\rm Joule}=\frac{1}{2\rho}{\bf B}\cdot{\nabla\times \bf E},
\end{equation}
such that magnetic energy is recycled as thermal energy after diffusion.

\subsection{Implementation of the boundary conditions}

The boundary conditions of Sect. \ref{s:bound} are incorporated
directly into the derivative operators at the boundaries and need to
account for sliding boundaries in the $x$ direction, periodic in $y$
and symmetric/antisymmetric in $z$ with additional density boundary
conditions also in the vertical direction. The $y$ boundary is fairly
trivial (the derivatives at the three points closest to each
$y$-boundary are defined such that they use points at the opposite end
of the box as well) however $x$ and $z$ boundaries are more complex
and are discussed below.

\subsubsection{$x$ boundaries}
 
The $x$-boundaries must take into account the sliding of boxes against
each other. We must therefore assume that the gridpoints in the
$y$-direction are not aligned between boxes so when calculating the
$x$-derivative at a boundary we must determine the amount of shear
that has occurred and then perform additional sixth order interpolation
to determine the values at the required position.

\begin{figure}
\caption{Calculations of derivatives at boundaries require
interpolation in the $y$-direction to determine values at intermediate
points. The figure shows one such $y$-interpolation required.
\label{f:xder}}
\end{figure}

Fig. \ref{f:xder} shows a typical situation where interpolation is
required to calculate the position at an intermediate point (
i.e. between gridpoints). This must be performed $3\times n_y\times
n_z$ times for each of the $x$ boundaries to be used with the sixth
order centred differences. The procedure also takes into account that
the points require for a particular interpolation may lie in adjacent
boxes in the $y$-direction.

\subsubsection{$z$ boundaries}

For the velocity, magnetic vector potential and energy we desire that
either the function value is equal to zero or the first derivative in
the $z$ direction is zero. These can be implemented using symmetric and
antisymmetric boundary conditions, respectively, to mimic points
outside the numerical domain. 

Fig. \ref{f:zder} shows that by specifying the points outside the
boundary such that, if we assume that the index of the boundary grid
point is $b$, then for a symmetric boundary condition
\begin{equation}
f_{b-i} = f_{b+i} \hspace{10mm}i=1...3,
\label{e:zbcs}
\end{equation}
results in a calculation that specifies
$\frac{\partial f_{z_{bc}}}{\partial z} = 0$. Similarly for an
antisymmetric boundary condition
of specifying that
\begin{equation}
f_{b-i} = - f_{b+i} \hspace{10mm}i=1...3,
\label{e:zbca}
\end{equation}
results in effectively setting $f_{z_{bc}} = 0$.

\begin{figure}
\caption{Calculations of derivatives at vertical boundaries can be
performed by specifying the function to be symmetric or antisymmetric
to produce $\frac{\partial f_{z_{bc}}}{\partial z} = 0$ and
$f_{z_{bc}} = 0$ respectively.
\label{f:zder}}
\end{figure}

The density boundaries are calculated from the hydrostatic equilibrium
condition Eq. (\ref{e:rhobc}). To implement this, we set
\begin{equation}
\ln \rho_{b+i}=\ln \rho_{b-i}+2 i \Delta z \frac{g}{(\gamma-1) e_b}.
\end{equation}

\subsection{Calculation of the time-step}

The time-step is limited by the Courant-Friedrichs-Lewy condition
\begin{equation}
\Delta t \leq \Delta t_c = \frac{\Delta x}{|{\bf{u}}| + c_s + v_a + |{\bf u}_0^{i=1}|},
\end{equation}

\noindent
where $\Delta_x$ is set to be the minimum mesh size over the three
directions, $v_a = (|{\bf B}|^2/\rho)^{1/2}$ is the Alfv\'en speed,
${\bf u}_0^{i=1}$ is the velocity of the underlying shear flow, and
$c_s=(\gamma p/ \rho)^{1/2}$ is the sound speed. This essentially states
that information must only be advanced a fraction of the mesh size for
each time-step. To guarantee the numerical stability, we choose a
safety factor $c_c \le 1$ (usually 0.3-0.5) so that the estimated
Courant time-step is
\begin{equation}
\Delta t = c_c \Delta t_c.
\label{e:cour}
\end{equation}

We also take into account diffusion when calculating the time
step. Stronger diffusion results in smaller time-steps and we take
into account the hyperdiffusion, shock viscosity and magnetic shock
dissipation. This condition can be expressed as
\begin{equation}
\Delta t_d = \frac{c_d \Delta x^2}{{\rm max}(\nu,\eta)}
\label{e:dif}
\end{equation}

\noindent for which both the shock and hyperdiffusive quantities of
$\nu$ and $\chi$ are included and where $c_d$ is an additional safety factor,
taken from empirical estimates to be of the order of 0.05. 

A radiative time-scale is included by taking the thermal conduction as
the relevant quantity. Taking a similar form the the above expression
we express this as
\begin{equation}
\Delta t_r = \frac{c_r \Delta x^2}{\chi^{\rm shk}+\chi^{\rm hyp}},
\label{e:rad}
\end{equation}

\noindent again using maximum values for $\chi$ and a safety factor of
$c_r=0.05$.

The final time-step is then derived from the minimum value of these
three time-scales and we find that this is adequate to ensure that the
code remains stable in all conditions.

\subsection{Comparison of time-stepping schemes}
\label{s:tscomp}

As mentioned earlier, we use an Adams-Bashforth-Moulton third order
predictor corrector scheme to advance the equations in time. We have
tested the performance of this compared a number of different methods
and found that it behaves favourably compared to them.

We have used the standard one-dimensional shock tube test (described in
more detail in Sect. \ref{s:hdriemann}) as a check on the accuracy
of the scheme since an exact analytical solution to this problem can
be found. This has been chosen as an adequate method of determining
the accuracy of a combination of different elements of the code
(namely the differencing operators in conjunction with the time
stepping). For reliable test results the resolution of the numerical
domain is varied and the time-step adjusted accordingly to more
realistically match the resolution (but fixed for
the duration of the test). In other words when the resolution is
doubled, the time-step is halved.

From the initial condition, the equations are advanced using a
constant time-step to a time of $t=0.256$. The error between the true
(analytical) and numerical results is calculated as a sum for density,
velocity and energy and averaged over the total number of
gridpoints. Hence the error, $\epsilon$, is given by

\begin{equation}
\epsilon = \frac{\Sigma_i (\rho-\rho_i)+\Sigma_i (u-u_i)+\Sigma_i
(e-e_i)}{n_x},
\end{equation}

\noindent where the quantities with subscript $i$ are the numerical
values and those without are the analytical values and $n_x$ is the
number of gridpoints.

As well as the third order Adams-Bashforth-Moulton method we have
performed the test on the second order Adams-Bashforth-Moulton method,
second and fourth order Runge-Kutta methods and the third order
predictor-corrector method of Hyman (Hyman \cite{Hyman79}). 

\begin{figure}[!ht]
\begin{center}
\caption{Errors incurred by the different schemes for the Sod tube
test using different resolutions and time-step sizes.  
\label{f:schmerrs}}
\end{center}
\end{figure}

The results of this test are shown in Fig. \ref{f:schmerrs}. Here we
see very clearly that as the resolution is increased and the time-step
shortened the higher order schemes produce smaller errors and in all
cases the errors from the Adams-Bashforth-Moulton third order scheme
are the smallest. It is for this reason that we have chosen this
method of advancing the equations. The exact algorithm used for this
scheme is given in Appendix \ref{a:time-steps}.


\begin{figure*}[!t]
\begin{center}
\caption{Data distribution over different numbers of dimensions. From
left to right these are one-dimensional, or slab distribution,
two-dimensional or column distribution and three-dimensional or block distribution.
\label{f:difdists}}
\end{center}
\end{figure*}

\section{Parallelisation}

\subsection{Methods}

 We chose to use High Performance Fortran (HPF) due to its simplicity
of implementing parallelisation methods and being ideally suited to
data parallelisation.  The finite difference model itself is ideal for
running on a number of processors supporting the Single Instruction
Multiple Data (SIMD) programming style where data is distributed onto
the local memory of each processor. The operations at a particular
grid-point are highly localised with data from only three
nearest-neighbour points being involved in any particular
operation. This results in a situation in which the data can be split
between a large number of processors, with communication between
processors only occurring at their common data boundaries. Hence, the
efficiency of the parallelisation should theoretically improve for
large data sets for which the boundary region size becomes small in
comparison to the size of the inner data region on a particular
processor (this fact is illustrated from the tests shown later in this
section). In effect, the data set on each processor can be operated on
virtually independently of all the other processors.

HPF directs the processor to distribute and align all the data
variables over a number of processors, effectively splitting the
domain into a number of smaller sub-domains. The communication calls
are automatically determined by the compiler. This provides a
particularly flexible approach to parallelisation since it can be
easily altered to match the relative number of gridpoints in different
dimensions.

\subsection{Parallelisation tests}

Many test have been performed to produce the most optimal
parallelisation results. Most of the tests have been performed on the
Cray T3E supercomputer at the Centre for Scientific Computing (CSC) in
Espoo, Finland. These include coding of derivative routines,
calculations of the shearing boundary conditions, ghostzone boundaries
and calculations of array operations. However, the most striking
results were determined for how best to distribute the data over a
number of processors. In theory the best distribution occurs for the
smallest surface area of boundaries between processors since
communication is at a minimum in this situation. However, our timings
indicate that this is not necessarily the best option with timings
widely varying between different distributions for the same
calculations.

To test the effectiveness of the different options available for
distributing data (one-dimensional slabs, two-dimensional columns or
three-dimensional blocks as illustrated in Fig. \ref{f:difdists}) we
use a simple test code that calculates derivatives 1000 times in all
three directions (assuming periodic boundaries) using a block of data
of equal dimensions ($63 \times 63 \times 63$). This data is
distributed over 8 processors chosen as it allows the processors to be
arranged in a cube when distributed in three dimensions for which the
communication should be minimised.  Results for this test are given in
Table \ref{t:dists} and shown in Fig. \ref{f:dists} for the total
times taken of all three derivatives. Following the notation of HPF
directives, distribution over a particular dimension is labeled as `B'
for `BLOCK' (data is distributed as a block onto a particular
processor in this direction) and `*' if distribution does not occur
along this particular direction.

\begin{table}[!t]
\caption{Comparisons of times of each derivative routine when varying
the distribution scheme. The total time taken for all three routines
is shown and this is plotted in Fig. \ref{f:dists}.
\label{t:dists}}
\begin{tabular}{ccccc}
\hline
Distribution & $x$-der. & $y$-der. & $z$-der. & total \\
\hline
\hline
(B,*,*) & 194.52 & 49.65 & 49.52 & 293.69 \\
(*,B,*) & 26.62 & 9.64 & 27.53 & 66.79 \\ 
(*,*,B) & 29.34 & 27.56 & 13.69 & 70.59 \\
(B,B,*) & 51.93 & 98.05 & 36.47 & 186.45 \\
(B,*,B) & 52.04 & 33.41 & 105.40 & 190.85 \\
(*,B,B) & 29.70 & 6.42 & 10.13 & 46.25 \\
(B,B,B) & 52.45 & 52.02 & 54.39 & 158.86 \\
\hline
\end{tabular}
\end{table}

\begin{figure}[!ht]
\caption{Comparisons of times to perform parallel calculations when
data is distributed over different directions. All distributions where
parallelisation has occurred over the $x$-direction show dramatically
decreased performance.
\label{f:dists}}
\end{figure}

From these tests we see that distributing in the $x$ direction
performs very poorly either alone or when distributed along with
others. It was also observed that speed up of codes in general is bad
when data is distributed along this direction.  This is probably
related to the Fortran column major ordering of arrays (i.e.
``first index changes fastest'') in memory which can lead to caching
problems.

The $y$ and $z$ distributions perform well and it is seen
that the distribution over both of these directions together performs
the best overall. This is presumably because the decreased surface
area of the boundaries between data distributed on the processors has
lead to less communication. This is seen by comparing the
$y$-derivative timing for the (*,B,*) distribution and the
$z$-derivative timing for the (*,*,B) with the same for the (*,B,B)
distribution in Table \ref{t:dists} in which both times are seen to
decrease when data has been spread more evenly in different directions.

\subsection{Shearing boundaries}

When incorporating shearing boundaries the results for distributing
data along the $y$-direction is less efficient since gridpoints at one
point on one side of the box in general need to communicate with
points at a totally different location on the other side of the box as
shown in the example of Fig. \ref{f:shearpara} where data has been
distributed over 4 processors in the $y$-direction.

\begin{figure}
\begin{center}
\caption{Communication required for shearing boundaries when
distribution occurs in the $y$-direction. In general,
the points required across the $x$-boundary for one particular
processor are not aligned on the same processor and are typically
stored in two processors located at any position within the computer.
\label{f:shearpara}}
\end{center}
\end{figure}

Here we see that due to the shearing, the calculations at one of the
$x$-boundaries on processors 1 require communication to acquire data on
processors 3 and 4. In general it could be any of the processors. This
is obviously much more expensive and complex than simply communicating
with points directly opposite the boundary. This in effect results in
poorer performance when data is distributed in this direction because
communication does not necessarily take place between neighbouring
processors. 

\begin{figure*}[!t]
\begin{center}
\caption{Different box dimensions for different applications of the
code. Each performs better under different parallel distributions
related to the grid dimensions. For example the galactic box performs
more efficiently for distribution in the  $z$-dimension due to the
grid resolution being weighted in this direction.
\label{f:boxes}}
\end{center}
\end{figure*}

\subsection{Performance of the full code}

From the above tests we see that the best performance of the parallel
code can be achieved by distributing data along the $y$ and $z$
directions. 

We also need to take into account the different model dimensions that
are likely to be used when determining which distribution method to
use. We therefore perform a number of tests on the speed up of the
code for different dimension models and different distributions. 

We chose to determine the speed up of the code by comparing timings of
the code when doubling the resolution and doubling the number of
processors. In an ideally parallelised code the real time of
communication should remain constant (however boundary conditions will
effect results). We also use this method rather than keeping a fixed
resolution and comparing timings when doubling the number of
processors due to the limiting fact that the Cray T3E contains only
128Mb of RAM of local memory on each processor. This implies naturally
that high resolution simulations can only be run on a large number of
processors and hence trends in speed-up are impossible to test. The
other alternative with this method is to use a low resolution
simulation that will run on a small number of processors. However,
this is an artificial test for two reasons: firstly the main aim is to run
high resolution simulations and secondly when distributed over a large
number of processors, communication time becomes artificially high since
only a small number of gridpoints in a particular direction lie on a
given processor. This test is designed to illustrate the essential
reasons behind parallelising the code; namely that through
parallelisation, high resolution simulations can be performed in
appreciable times.

The test uses all parts of the code including magnetic terms, shearing
and numerical diffusion. The test is not aimed at solving a physical
problem but the data is initialised as it could be for a typical
galactic run with a large scale azimuthal magnetic field. The code is
then timed for performing 30 time-steps.

Three tests are displayed here showing the results for different
distributions of data corresponding to different shapes of boxes that
are commonly used in different astrophysical systems as shown in
Fig. \ref{f:boxes}. For the galactic setup, distribution in the
$z$-direction results in the best distribution of data amongst
processors while for the accretion disk and stellar applications, $y$
and $y$-$z$ distributions are best. Hence grid resolutions and
distributions of data are chosen to match each of these situations as
closely as possibly. For all the tests performed we have
roughly an equal number of gridpoints per processor making comparisons
between distributions possible also.

For ideal parallelisation, the real time of calculation should remain
constant when doubling the number of processors and doubling the grid
resolution. We therefore measure performance on real time per
grid-point. These are then scaled as a percentage of the time taken on
two processors.

Table \ref{t:times1} shows the times for distributing the data in the
vertical direction for a grid resolution that is biased towards the
vertical. Fig. \ref{f:times1} shows plots of the relative times and
speed up of the code. Tables \ref{t:times3} and \ref{t:times2} along
with Figs. \ref{f:times3} and \ref{f:times2} respectively show the
results for the alternative types of box dimensions and data
distributions.

\begin{table}[!ht]
\caption{Times for distributing the data (*,*,BLOCK) typical for a
galactic simulation.
\label{t:times1}}
\begin{tabular}{cccccc}
\hline
No. of  & Grid       & Real time & Relative \\
procs.  & resolution & (s)       & speed up \\
\hline
\hline
2 & $31 \times 31 \times 255$ & 742.61  & 2.00 \\
4 & $31 \times 63 \times 255$ & 764.51  & 3.95 \\
8 & $63 \times 63  \times 255$ & 711.29 & 8.62 \\
16 & $63 \times 63 \times 511$ & 708.45 & 17.40 \\
32 & $63 \times 127 \times 511$ & 738.99 & 33.56 \\
64 & $127 \times 127 \times 511$ & 734.29 & 68.26 \\
128 & $127 \times 127 \times 1023$ & 732.26 & 136.99 \\
\hline
\end{tabular}
\end{table}

\begin{figure}[!ht]
\begin{center}
\caption{Performance of the code starting from $31\times 31\times 255$
to $127 \times 127 \times 1023$ from 2 to 128 processors distributing
the data as (*,*,BLOCK). See Table \ref{t:times1}  for intermediate sizes and
actual timings
\label{f:times1}}
\end{center}
\end{figure}

\begin{table}[!ht]
\caption{Times for distributing the data (*,BLOCK,*) typical for an
accretion disk simulation.
\label{t:times3}}
\begin{tabular}{cccccc}
\hline
No. of  & Grid       & Real time & Relative \\
procs.  & resolution & (s)       & speed up \\
\hline
\hline
2 & $31 \times 255 \times 31$ & 879.23 & 2.00 \\
4 & $31 \times 255 \times 63$ & 905.84 & 3.94 \\
8 & $63 \times 255 \times 63$ & 808.41 & 8.97 \\
16 & $63 \times 511 \times 63$ & 802.86 & 18.18 \\
32 & $63 \times 511 \times 127$ & 862.69 & 34.01 \\
64 & $127 \times 511 \times 127$ & 870.33 & 68.03 \\
128 & $127 \times 1023 \times 127$ & 898.92 & 131.58 \\
\hline
\end{tabular}
\end{table}

\begin{figure}[!ht]
\begin{center}
\caption{Performance of the code starting from $31\times 255\times 31$
to $127 \times 1023 \times 127$ from 2 to 128 processors distributing
the data as (*,BLOCK,*). See Table \ref{t:times3} for intermediate sizes and
actual timings
\label{f:times3}}
\end{center}
\end{figure}

\begin{table}[!ht]
\caption{Times for distributing the data (*,BLOCK,BLOCK) typical for a
stellar simulation.
\label{t:times2}}
\begin{tabular}{cccccc}
\hline
No. of  & Grid       & Real time & Relative \\
procs.  & resolution & (s)       & speed up \\
\hline
\hline
2 & $63 \times 63 \times 63$ & 811.98 & 2.00 \\
4 & $63 \times 63 \times 127$ & 824.25 & 3.96 \\
8 & $63 \times 127 \times 127$ & 827.86 & 7.96 \\
16 & $127 \times 127 \times 127$ & 772.57 & 17.24 \\
32 & $127 \times 127 \times 255$ & 778.36 & 34.30 \\
64 & $127 \times 255 \times 255$ & 796.55 & 68.5 \\
128 & $255 \times 255 \times 255$ & 792.46 & 136.02 \\
\hline
\end{tabular}
\end{table}

\begin{figure}[!ht]
\begin{center}
\caption{Performance of the code starting from $63\times 63\times 63$
to $255 \times 255 \times 255$ from 2 to 128 processors distributing
the data as (*,BLOCK,BLOCK). See Table \ref{t:times2} for intermediate sizes and
actual timings
\label{f:times2}}
\end{center}
\end{figure}

We see that out of the three tests, the times for the vertical
distribution are the best. This is because the shearing boundaries are
not affected by the parallelisation. Data that is required by one
particular point on an $x$-boundary always lies on the same processor
therefore expensive communication does not occur. We also note, that
in general, as the number of processors increases (along with the grid
resolution), the performance appears to improve which is seen from the
graphs where the speed up is above optimal. This is due to
the percentage of time taken up in calculating the boundary conditions
diminishing as the overall size of the model increases. All the
results show that the code has good speed up when doubling the
resolution and doubling the number of processors with virtually linear
speed up in all cases.

\section{Tests}

\begin{figure*}
\caption{Results for the weak Riemann shock-tube test where the
density on the left is initially set to $\rho=1.0$ at $t=0.245$ for
255 gridpoints. Figures plotted are for density, pressure, velocity
and energy. The dashed line shows the analytical solution.
\label{f:rw}}
\end{figure*}

The code has been tested against a number of standard hydrodynamical
and magneto-hydrodynamical tests in 1D, 2D and 3D. These tests have
been designed to test individual properties of the code as well as all
components of the code working together. Some of these tests are
`standard' tests of fluid models and others have been incorporated to
compare the results of our method against the existing analytical
solutions and other numerical models.

Most of the hydrodynamical tests are designed to determine the
effectiveness of the numerical viscosities in stabilising the code,
removing unphysical features and capturing the essential physics of a
particular problem. They are also aimed at determining the ability of
the Cartesian grid to reproduce spherical features. For all tests we
use $c_{\rm shk}=2.0$ and $c_{\rm hyp}=0.05$ for the coefficients of
the shock and hyperviscosities. The Prandtl number and magnetic
Prandtl numbers are equally set to unity.

Using the MHD test suite of Stone et al (\cite{StoneHEN92}) and Stone \&
Norman (\cite{StoneN92b}) as a basis for the magneto-hydrodynamic tests,
we perform a number of identical tests again using previously
published results for comparison with our numerical method. These
tests are designed to examine the stability of the fluid and magnetic
field evolution and more specifically to check that the
characteristics of the MHD flow are correct, specifically the
propagation of MHD waves.

These tests have allowed us to determine the effectiveness of the
algorithms used, as well as to show their weaknesses. All tests have
been performed using resolutions comparable to those expected in
`real' simulations and hence act as a gauge on the accuracy one can
expect from future calculations.

\subsection{Riemann shock tube test}
\label {s:hdriemann}

\begin{figure*}
\caption{Results for the strong Riemann shock-tube test where the
density on the left is initially set to $\rho=10.0$ at $t=0.150$ for
255 gridpoints. Figures plotted are the natural logarithm of density,
pressure, velocity and energy. The analytical solution is plotted as a
dashed line.
\label{f:rs}}
\end{figure*}

The Riemann shock tube test (Sod \cite{Sod78}) has been used by many
authors as a test of numerical algorithms. All components of the
hydrodynamical code are used including numerical viscosities. This
test determines the ability of the code to capture shocks, formed from
discontinuities in the fluid properties, and therefore it particularly
is a direct test of the shock viscosity employed.

Starting from an initial discontinuity in the density and energy, the
fluid forms a shock front and a rarefraction wave. For this test an
exact analytical solution can be found (e.g. Hawley et
al. \cite{HawleySW84}), which provides an ideal situation for
determining the accuracy of the code including the propagation speed
of the waves, the jump conditions at the shock front, and the ability
of the code in stabilising discontinuities within the fluid.

We perform two test cases; one using the standard setup described by
Sod (\cite{Sod78}) and another for which the shock is much
stronger. This second test is included as a more accurate measure of
the ability of the code to cope with more violent physical features,
and as a more extreme test of the numerics.

Both tests are calculated in one dimension (we use the z-dimension
with closed boundaries) with a resolution of $n_z=255$ with
$z=\{0,1\}$. The gas has a ratio of specific heats of $\gamma=1.4$
with zero initial velocity. The density and energy, and therefore
pressure, are discontinuous at $z=0.5$. For the weak (standard) shock
tube test, we have on the left $\rho_l=1.0$ and with the strong shock
tube test this is $\rho_l=10.0$. All other variables are the same for
the two tests with $P_l=1.0$, $\rho_r=0.125$ and $P_r=0.1$.

Fig. \ref{f:rw} shows the evolution of the variables for the weak
shock at $t=0.245$ compared to the analytical solution, shown as a
dashed line. Many features of the fluid properties have been captured
well by the calculation. The position of the shock front, the contact
discontinuity and the rarefraction wave are all correct and the
magnitudes of the fluid properties in each region is correct. The main
deviations from the analytical curve occur at the contact
discontinuity and the shock front. At the shock front the shock is
captured well within four gridpoints however a slight under-shoot
occurs for the energy which is quickly corrected within three
gridpoints. The model has not successfully reproduced the shape of the
contact discontinuity for energy or density, with both variables being
smoothed. This feature appears to be common amongst many different
numerical schemes as shown in the figures by Sod (\cite{Sod78}) and
Stone \& Norman (\cite{StoneN92}). The discontinuities in the gradient
of the rarefraction wave have been captured well by the scheme,
although some smoothing is inevitable.

Fig. \ref{f:rs} shows the equivalent variables at $t=0.150$ for the
strong initial discontinuity, with density plotted as a natural
logarithm. This, being a more rigorous test of the numerics, shows
more unphysical features than the weak shock. However, the different
regions of the flow are very clear and follow well from the true
values - the main difference being the ability of the code to retain
the strong shock features. The positions of the boundaries between the
different regions all agree very well to the analytical solution,
however it is noted that the shock front appears to have moved
fractionally further. Small scale oscillations in the velocity are
seen to be generated behind the shock, which is natural of a scheme of
this type, however, the hyperdiffusion has minimised the magnitude of
these. Again, the contact discontinuity is smeared by the simulation,
and the plateau of maximum energy is not resolved, but does not
deviate far from the true value. The shock front itself is still well
resolved and retains the essential physical characteristics showing
that the shock viscosity is implemented correctly. In this case, the
under-shoot is smaller than for the weak shock. The rarefraction wave
still clearly remains close to the analytical solution and the
discontinuous gradients are well represented. Considering the strong
initial conditions presented to the flow, we feel that the numerical
solution presents a good fit to the analytical one.

\subsection{Blast waves}

\begin{figure*}
\caption{Lefthand panel shows spherically symmetric expansion of an
adiabatic blast wave, showing grey-scale representation of logarithmic
density at the horizontal mid-plane of a 3D simulation using 127$^3$
gridpoints at 22 Myrs. On the right we show the 1D profiles of density
and velocity during the shock stage at 3 Myrs for the resolution 255
plotted on the analytical Sedov-Taylor solution.
\label{f:sedov_profiles}}
\end{figure*}

\begin{figure*}
\caption{Position of the expanding remnants for different setups. The
leftmost panel shows the expansion of adiabatic and radiative remnants
in three dimensions. The panel on the right shows the expansion of
adiabatic remnant in a uniform azimuthal magnetic field. The dashed
line in both figures shows the Sedov-Taylor expansion law with the
slope 0.4. For the magnetic case the expansion of the remnant in the
azimuthal direction is shown with a solid line, while in the
$x$-direction it is dot-dashed. The magneto-sonic wave moving in the
$x$-direction is represented by the dotted line.
\label{f:expansion3d}}
\end{figure*}

\begin{figure*}
\begin{minipage}{0.5\linewidth}
\hspace{-0.7cm}
\end{minipage}
\hspace{1.cm}
\begin{minipage}{0.5\linewidth}
\end{minipage}
\caption{Interaction of blastwave with large-scale azimuthal magnetic
field. The lefthand panel shows a 2D slice through the horizontal
midplane with density shown in grey-scale with velocity field plotted
on top with arrows. The righthand panel shows the voxel projection of
the density with the perturbed magnetic field lines plotted on
top. The expansion of the shock front is seen to be restricted to the
azimuthal direction while the magneto-sonic wave perturbs the density
in the poloidal directions.
\label{f:magblast}}
\end{figure*}

The next set of tests is aimed at testing the capability of the code
to model spherical features with a Cartesian grid. We perform three
3D tests of strong blast waves: adiabatic, radiative, and with a strong
imposed magnetic field.

First we follow the evolution of an adiabatic shock created by an
instant injection of purely thermal energy, monitoring its shape,
radius and expansion velocity, and comparing it to the analytical
Sedov-Taylor solution (Taylor \cite{Taylor50}; Sedov
\cite{Sedov59}). The explosion is initialised by adding $10^{51}$ ergs
of thermal energy, roughly corresponding to a realistic SN explosion
(e.g. Heiles \cite{Heiles87}), in a single grid-point in the middle of
the computational volume sized 1 kpc$^{3}$. The number of gridpoints
used in the test is 127$^3$. The surrounding ISM has a uniform
density of 1.0 cm$^{-3}$ and temperature $10^4$K without any magnetic
field. No heating or cooling terms are applied to the energy equation
(\ref{e:ene}). According to the Sedov-Taylor solution, based on
similarity analysis, the shock front produced by a strong explosion in
a uniform medium (in a three-dimensional volume) moves through it so
that its radius as function of time is
\begin{equation}
R(t)={E}{\rho_0}^{2/5} t^{1/5},
\end{equation}
where $E$ is the explosion energy and $\rho_0$ is the density of the
surrounding ISM. Firstly we compare the expansion of the simulated
blastwave to the Sedov-Taylor solution by plotting its radius versus
time in a logarithmic scale in Fig. \ref{f:expansion3d}. If the
blastwave is to follow the Sedov-Taylor solution, there should be a
powerlaw with the slope 2/5 visible in this figure, and in addition
the simulated radius should fall on top of the dashed line
representing the Sedov-Taylor solution for the selected $E$ and
$\rho_0$. Initially the blastwave is observed to expand faster than
the Sedov-Taylor theory predicts, which is due the free expansion
phase during which the explosion front has not yet swept up enough
mass to form a shock front and therefore expands freely. In reality SN
explosions, and also the free-expansion phase, occur at much smaller
scales than the grid resolution of numerical models, so the ``late''
free expansion phase observed here is an artifact from the finite
resolution. After the first few tens of thousands of years of
evolution, when the shock has formed, the remnant starts following the
Sedov solution rather closely, and only after 4 million years the
expansion starts slightly deviating from that powerlaw. At that point
the expansion velocity of the remnant has become comparable to the
local sound speed, when the shock dies out, and the remnant starts
dissolving to the surrounding ISM. The previously thin shock front
forms a thicker shell, and matter is flowing in and out as the shell
relaxes. In Fig. \ref{f:sedov_profiles} we also show the shape of the
blast wave in the horizontal ($xy$) plane, and the 1D profiles of
density and velocity on top of the ones calculated from the
Sedov-Taylor solution. Throughout the calculation the blast wave
remains spherical as can be seen from the left panel of
Fig. \ref{f:sedov_profiles}, which shows the remnant at later stages
(22 Myrs). During the shock stage the profiles of the physical
quantities resemble quite well the analytical profiles, although the
jump conditions are not quite satisfied at the shock front (on the
right in Fig. \ref{f:sedov_profiles}).

The Sedov solution serves as a good approximation for a SN explosion
when the radiative losses are negligible, which is true roughly during
the first 10$^5$ years of its evolution (e.g. Shu \cite{Shu92}). When
the radiative losses become significant, they change the expansion
characteristics of the remnant, as discussed e.g. by Ostriker \& McKee
\cite{OM88}. We investigate a radiative remnant with the same setup as
used for the adiabatic case, but adopting a cooling function derived
by Dalgarno \& McCray (\cite{dalgarnomc72}) and Raymond et al.
(\cite{RaymondCS76}) that has been previously used in several ISM models
(e.g. Rosen \& Bregman \cite{RosenB95} and V\'azquez-Semadeni et
al. \cite{VazquezP95}), which is implemented as a sink term in the
energy equation (\ref{e:ene}) so that
\begin{equation}
\frac{\partial e}{\partial t}= \ldots -\rho \Lambda,
\end{equation}
where $\Lambda=\Lambda_i T^{\beta_i}$, $T_i \leq T \le T_{i+1}$, is
the piece-wise cooling function described in Table
\ref{t:coolingtable}. The position of the radiative shock front is
shown in Fig. \ref{f:expansion3d}, whereby it can be seen that
during the first tens of thousands of years the expansion of the blast
wave follows rather closely to the adiabatic case, but after
approximately $10^5$ years the two expansion curves start deviating
from each other, indicating that the radiative losses have become
significant. After that time the remnant shows powerlaw expansion as
in the adiabatic case, but the slope of the line is considerably less,
about $0.29$, which is close to the slope 1/4 reported in Ostriker \&
McKee (\cite{OM88}) for a radiative blastwave in homogeneous medium. Due
to the energy loss via radiation, the expansion velocity drops much
faster than in the adiabatic case, being comparable to the sound speed
already at one million years.

\begin{figure*}
\caption{Results for the interacting blastwaves, taken in
one-dimension. The points show the resolution of 511 gridpoints and
the solid line is for 8191 gridpoints. Plots of velocity and density
are shown at three different times ($t=0.010$, $t=0.028$ and
$t=0.038$). The lower resolution case closely follows the evolution of
the more accurate high resolution simulation however strong peaks are
damped.
\label{f:tb1d}}
\end{figure*}

\begin{figure*}
\begin{center}
\end{center}
\caption{The two-dimensional evolution of interacting blastwaves
generated from two equally strong point explosions (corresponding to a
typical SNe of $10^{51}$ ergs) in a uniform medium with no magnetic
field shown at three different times ($t=5$Myrs, $t=12$Myrs and
$t=34$Myrs. The collision occurred at $t=1$Myr. Grey-scale and black
contours shows logarithmic density with velocity field plotted with
white arrows on top. The resolution is fixed at 127 gridpoints per
kpc.
\label{f:tb2d}}
\end{figure*}

\begin{figure*}
\caption{Advection of pulse of magnetic field in the $z$
direction. The upper two panels show the inital condition of the test
with the left plot showing the radial magnetic field and on the right
the azimuthal current density. The deviation from a square wave is due
to the use of the magnetic potential. The final state is shown in the
lower two panels The resolution is $n_z=255$ and for a perfect
initially square pulse, the edges would be located at $z=255$ and
$z=305$.
\label{f:mhdadv}}
\end{figure*}

\begin{figure*}
\caption{Propagation of shear Alfv\'en wave in a fluid threaded by a
vertical magnetic field.  The upper panel is for an initially
stationary fluid and the lower for a fluid with initial velocity of
$u_z=1.5$. For the stationary fluid, the wave is generated by a
perturbation in velocity that was initially located between $z=1$ and
$z=2$ whereas for the non-stationary case it is located between $z=2$
and $z=3$. Both cases show that square pulses of magnetic field are
generated, propagating perpendicular to the applied magnetic field at
the correct velocity. The resolution is set to $n_z=255$.
\label{f:mhdshear}}
\end{figure*}

\begin{table}[t]
\caption{The cooling function used for investigating radiative
blastwaves.
\label{t:coolingtable}}
\begin{tabular}{ccc}
\hline
$T_i$ [K] &$\Lambda_i$ [erg s$^{-1}$ g$^{-2}$ cm$^3$] & $\beta_i$ \\
\hline
\hline
100  &1.14 $\times 10^{15}$ &2.000 \\
2000 &5.08 $\times 10^{16}$ &1.500 \\
8000 &2.35 $\times 10^{11}$ &2.867 \\
$10^5$ &9.03 $\times 10^{28}$ &-0.650 \\
\hline
\end{tabular}
\end{table}

Finally, we investigate adiabatic blast waves in the presence of a
strong azimuthal magnetic field. The setup is identical to the
adiabatic case, but now we impose an azimuthal magnetic field of the
strength 5$\mu$G at each time-step, and follow the position of the
blast wave in the direction along the field lines ($y$-direction), and
perpendicular to them ($x-$direction), which curves are shown in
Fig. \ref{f:expansion3d}. The shape of the blast wave is shown as
density contours over-plotted with velocity field vectors at the
horizontal mid-plane in the left-hand panel of
Fig. \ref{f:magblast}. In the right-hand side of Fig. \ref{f:magblast}
we show a voxel projection of the 3D density field with perturbed
magnetic field lines. All these figures illustrate how the magnetic
field can severely affect the expansion. Along the magnetic field
lines the blast wave expands in a normal fashion and develops a strong
shock front. However due to magnetic tension of the field lines, this
does not occur along the $x$ direction. We also see formation of a
magnetosonic wave, which propagates perpendicular to the field lines,
forming a weak spherical perturbation, slightly pinched at the
poles. The shape of the perturbation is similar to the one described
by Ferri\'ere et al. (\cite{FerriereMMZ91}). In the $x$-direction the
expansion velocity of the blast wave is heavily damped, and most of
the expanding motion occurs in the $y$-direction, as seen e.g. in the
simulations of Tomisaka \cite{Tomisaka98}). In
Fig. \ref{f:expansion3d} the expansion in the $y$-direction is seen to
roughly follow the Sedov-Taylor law, but the $x$-direction strongly
deviates from it. The expansion of the magnetosonic wave is much
faster than the non-inhibited blast wave, since it is moving with the
Alfv\'en velocity 14 km s$^{-1}$, which after approximately 4 Myrs
becomes faster than the blast wave itself. In three dimensions the
spherical blast wave is unrecognisable. Complex features have been
formed, where the magnetosonic wave has created a lemon-shaped weak
density perturbation within which the blast wave has produced a cavity
elongated in the $y$-direction.

\subsection{Interacting blast waves}

We perform two tests, in one and two dimensions, to show the ability
of the code to deal with interacting blast waves. The first follows
the one-dimensional colliding blast wave test presented by Woodward \&
Colella (\cite{WoodwardC84}) who performed this test on various
algorithms comparing a very high-resolution case to lower-resolution
ones. We perform a similar test, comparing a high-resolution
calculation to a moderate resolution case. One reason for this test is
to compare the high-resolution case to the Woodward \& Colella case as
a check on shock velocities and density profiles, and a second reason
is to see how the code adapts with different resolutions. If the code
scales well (and in particular the numerical viscosities) between
different resolutions then one would expect the shock positions and
profiles to be in close agreement.

The test setup follows that of Woodward \& Colella
(\cite{WoodwardC84}). The numerical domain is in the vertical
direction, again for reasons of boundary conditions, with $z=\{0,1\}$
and the ratio of specific heats set to be $\gamma=1.4$. Velocity is
initially at zero with density everywhere set to be $\rho=1.0$. Two
shock waves are generated by setting two discontinuities in the
pressure. For $z=\{0.0,0.1\}$ we set $P=1000$, $z=\{0.1,0.9\}$
$P=0.01$ and for $z=\{0.9,1.0\}$ we have $P=100$, hence two blast
waves of different magnitudes are generated moving towards each other.

Since no exact analytical solution exists, we perform a very
high-resolution test calculation to obtain profiles which can be
considered to be highly accurate. For this we set $n_z=8191$
which allows the sharp density peak at the time of collision to be
well resolved. For the moderate resolution, we set $n_z=511$.

Fig. \ref{f:tb1d} shows the evolution of the fluid at three different
times, each of which can be compared to the figures shown by Woodward
\& Colella (\cite{WoodwardC84}). The very high-resolution calculation
is shown as a dashed line with the moderate resolution plotted on top
as diamonds. Evident from all the figures is that the shocks travel at
identical speeds for both resolutions, an important factor when using
the code at different resolutions. One also observes that the shapes
of the curves are almost exactly the same. At $t=0.028$ we see a
slight deviation in the velocity behind the shock front, but in all
other aspects and at other times the different resolutions appear to
be virtually identical. The lower resolution run obviously cannot
resolve very small features, and this is evident at the time of
collision when the density spike is smaller, but is in the correct
position. At the final stage the density minima and maxima are again
smoothed as a consequence of the lower resolution, but retain the same
shape as the very high-resolution case. Overall, we feel that the
lower resolution simulation compares very well to the high resolution
case, and compared to the original Woodward and Colella figures of
lower resolution calculations (which are however for 1200 grid zones)
performs very well.

The second test we perform is a two-dimensional test on colliding
spherical remnants produced by two equally strong explosions in a
uniform medium, which is the configuration discussed e.g. by Courant
\& Friedrichs (\cite{CourantF48}), and studied also with numerical
models e.g. by Yoshioka \& Ikeuchi (\cite{YoshiokaI90}) and Voinovich
\& Chernin (\cite{VoinovichC95}). We again initialise the two
explosions as thermal energy releases in a single grid-point
corresponding to a SN energy of $10^{51}$ ergs, located 0.4 kpc apart
from each other. The remnants collide after 1 Myr, having equal
expansion velocities, and instantly after the collision a reflected
shock front is formed propagating back into the hot and sparse
interiors, as seen in the top left panel of Fig. \ref{f:tb2d}. At the
location of collision, a tangential line forms, along which the radial
flow stops persisting throughout the simulation, and thereby seen in
all the panels of Fig. \ref{f:tb2d}. Soon afterwards a new shock
front, denoted as the Mach front by Courant \& Friedrichs
(\cite{CourantF48}) traveling along this line appears, with the
velocity exceeding the expansion of the unperturbed remnants. This
configuration seen in the simulation at this stage closely resembles
the classical Courant-Friedrichs picture. The top right panel of
Fig. \ref{f:tb2d} shows that after about 10 Myrs the Mach front has
propagated outwards and increased its surface area, so that the whole
systems starts becoming spherical on the outside, even more pronounced
in the lower panel at 34 Myrs. At the same time the curved reflected
shock fronts expand in the hot sparse interior having largest
expansion velocities in the direction of the smallest density with
vortical motions occurring leading to kidney-shaped structures. The
appearance of the flow is in good agreement with other numerical
simulations, such as the detailed calculation presented by Voinovich
\& Chernin (\cite{VoinovichC95}) and also with the results of Yoshioka
\& Ikeuchi (\cite{YoshiokaI90}).

\subsection{Magnetic advection test}

We now perform an advection test of a pulse of magnetic field. The
test is initialised such that a pulse of magnetic field, which is as
close as possible to a square wave, is advected at constant velocity
in one direction. This is perhaps the least relevant test for the
numerical scheme as it requires that most of the PDEs are disabled. In
other words, the square pulse has no back-reaction on the fluid in any
sense. The test is essentially used here to study the numerical
diffusion of the wave (since numerical diffusion acts all the time and
quite strongly where fluid properties vary rapidly between gridpoints)
ensuring that numerical instabilities are quenched and that the
evolution of the current that is generated at the leading and trailing
edges of the pulse is correct.

The test is performed in one dimension along the vertical axis. The
use of the $z$-direction is different to Stone \& Norman
(\cite{StoneN92b}) and arises from the fact that boundary conditions
in the vertical direction mean that setting up tests such as this (and
subsequent tests shown below) is simplified. Discontinuous values of
the magnetic vector potential in periodic directions result in
propagation of waves waves from the boundaries. This change in
direction also results in the use of $B_x$ which in effect means that
the test should be otherwise identical. However, the use of the
magnetic vector potential causes a number of difficulties when
initialising the wave and obtaining a perfect square wave is
impossible since discontinuities in the magnetic vector potential lead
to ringing occurring at the corners of the wave. We therefore start
the test from a wave in which the edges of the pulse are spread over
three grid-zones rather than just one, which is already slightly
smoother than that of Stone \& Norman. However we as closely as
possible follow their setup. The total width of the pulse is over 54
gridpoints, with the upper values covering 48. The initial shape of
the field and current is shown in the upper panels of
Fig. \ref{f:mhdadv}. The pulse is then advected over 250 gridpoints
for which $\Delta z = 1$ with a velocity of $u_z=1$.

The final state of the advected magnetic field and current is shown in
lower two panels of Fig. \ref{f:mhdadv}. As expected from the use of
the diffusive elements of the model, the edge of the pulse has now
been smeared from the initial three grid-zones to approximately
10. The current density is therefore similarly smeared. As quoted by
Stone \& Norman, the use of the vector potential can lead to the
production of non-monotonic currents. Indeed, it is seen that the
trailing edges are non-monotonic. However, no sign reversal is seen
and the deviations from monotonicity are very small in comparison to
the maxima of the current.

\subsection{Propagation of shear Alfv\'en waves}

The next test we perform is to propagate Alfv\'en waves initially
generated from a shear flow. The test is again set up identically
to that of Stone \& Norman (\cite{StoneN92b}) except for the use of the
$z$ direction again for reasons stated above.

The test is intialised by threading the fluid with a vertical magnetic
field, $B_z$, and a small perturbation to $u_y$ is added to a small
region of the width one, the extent of the domain being 15. In terms
of dimensionless units, $\rho=1$, $B_z=1$ and $u_y=0.001$ in the
perturbed region. As with Stone \& Norman, two tests are performed:
one for which $u_z=0$ and the perturbed velocity is between the
regions of $z=1$ and $z=2$ and one for $u_z=1.5$ with the perturbation
occurring between $z=2$ and $z=3$. The shearing motion then generates
square Alfv\'en waves propagating at an Alfv\'en velocity of $v_A=\pm
1$ (with $\mu_0=1$). For the first case of this test, these propagate
with and equal velocity in each direction ($u=\pm 1$) and for the
second case one has an overall velocity of $u=u_z+v_A=2.5$ and the
other $u=u_z-v_A=0.5$. Both calculations have a resolution of
$n_z=255$.

Fig. \ref{f:mhdshear} shows the final state of the azimuthal
velocity, $u_y$, and sheared magnetic field, $B_y$, for the two
cases. For the first case the plots are shown after $t=0.8$. The
waves have clearly traveled the correct distance of $z=0.8$
corresponding to $u=1$ and have propagate evenly in both directions away
from the initial sheared region. The second figure is shown after a
time of $t=1.0$, when the two waves have traveled distances of $z=2.5$
and $z=0.5$ corresponding to the right and left waves
respectively. Again these correspond correctly to the overall
velocities of $u=2.5$ and $u=0.5$. Also, the widths of the square
pulses are almost identical to the initial width of the sheared region
in both cases.

It can be seen that in the second case the diffusion has acted more
strongly. This is due to the magnitude of the hyperdiffusion depending
upon both Alfv\'en velocity and fluid velocity and hence being greater
for the case in which $u_z=1.5$.

\subsection{Magnetic breaking of an aligned rotator}

\begin{figure*}
\caption{Results of the magnetically braked aligned rotator showing
the resulting azimuthal velocity and magnetic field at $t=13$. The fluid
is threaded by a vertical magnetic field and the fluid density remains
fixed with $\rho=10$ between $z=0$ and $z=1$ and $\rho=1.0$ for the
remaining domain. The high density region is initially given a
perturbing azimuthal velocity which generates Alfv\'en waves moving
both inwards and outwards. The dashed line shows the analytical
solution. The resolution is set to 301 gridpoints.
\label{f:mhdrot}}
\end{figure*}

\begin{figure*}[!ht]
\begin{center}
\end{center}
\caption{Results for MHD Riemann problem. The setup follows exactly
that of the hydrodynamic counterpart with an additional discontinuous
vertical magnetic field, aligned with the other fluid
discontinuities. The subsequent evolution of various kinds of waves is
shown at $t=80$ for density, pressure, vertical velocity, radial
velocity and radial magnetic field. The grid resolution is set to 801
gridpoints. 
\label{f:mhdrmn}}
\end{figure*}

The third test is the magnetic breaking of an aligned rotator. This is
virtually identical to the previous test in which the fluid is
threaded by a magnetic field and a region of the fluid is then
perturbed azimuthally. For this test however, there is a density
contrast between the perturbed region and the steady region resulting
in a setup which mimics a disk of high density surrounded by a low
density medium. The perturbation generates Alfv\'en waves propagating
away from the disk and also into the disk thus accelerating the fluid
in the surrounding medium and decelerating the disk. Subsequent
partial reflections from the surface of the disk further complicate the
system, each being both transmitted as a lower amplitude waves into
the medium and back into the disk. A more comprehensive discussion of
the model is given by Stone \& Norman (\cite{StoneN92b}).

Again, using identical parameters to Stone \& Norman, the disk is of
density $\rho_d=10.0$ and the surrounding medium is of
$\rho_m=1.0$. The test again is in the $z$-direction with 300
grid-zones. The disk is located in the region $z < 1$ and the low
density medium extends from this point to $z=15$. The vertical
magnetic field has a strength of $B_z=1$. Setting $u_y=0.001$ in the
region $z<1$ and zero elsewhere, we are able to make comparisons to
the analytical results of Mouschovias \& Paleologou
(\cite{MousP80}). The profiles of the resulting sheared magnetic
field, $B_y$, and azimuthal velocity, $u_y$, are independent of the
initial velocity (varying only in magnitude).

Fig. \ref{f:mhdrot} shows the final state of the relevant quantities
after $t=13$. The dashed line shows the analytical values for
comparison. The waves, which are generated from the initial shear and
subsequent reflections at the disk surface, closely match the profiles
of the analytical values but are smoothed due to the inherent
diffusion of the scheme. However, propagation speed and magnitudes
are correct.

\subsection{MHD Riemann shock tube test}

The next test is the MHD equivalent to the Riemann shock tube test
(Sod \cite{Sod78}) . This has been discussed in detail for a number of
numerical schemes by Brio \& Wu (\cite{BrioW88}) and used as a
subsequent test by Stone \& Norman (\cite{StoneN92b}). This tests uses
all elements of the code to evolve a fluid that is initiated with a
discontinuous pressure and magnetic field at the midpoint with an
additional component of the magnetic field along the direction of
motion. Unlike the standard shock tube test, no known analytical
solution exists for the subsequent evolution of the fluid and magnetic
field. Hence we make comparisons to the previously published works
mentioned above.

The hydrodynamical parts of the test are initialised exactly as in the
hydrodynamical counterpart. The domain size of the problem is however
larger (800 grid-zones with $\Delta z=1$ to match the published results
of Brio \& Wu (\cite{BrioW88}) and Stone \& Norman
(\cite{StoneN92b}). We again use the $z$ direction due to the boundary
conditions and change the direction of the discontinuous magnetic
field accordingly. The discontinuity is at the centre of the
domain. Fluid to the left takes physical values of $\rho_l=1.0$, $P_l=1.0$
and $(B_x)_l=1.0$. and to the right takes $\rho_r=0.1$, $P_r=0.1$ and
$(B_x)_r=-1.0$. An additional magnetic field of $B_z=0.75$ is constant
throughout the domain. As in the previously published MHD Riemann
shock tube tests, we take $\gamma=2$. As with the magnetic advection
test, the use of the magnetic vector potential produces problems when setting
up discontinuous magnetic fields. The initial
discontinuity is therefore spread over three grid-zones. However we feel
that, with the use of magnetic shock viscosities and hyperdiffusion,
this small initial difference has negligible effects on the resulting
evolution.

Unlike the hydrodynamical case, the MHD shock tube test generates many
kinds of waves as well as the shock and rarefraction wave. As quoted
by Brio \& Wu, the fluid can contain a compound wave which consists of
a shock wave attached to a rarefraction wave of the same family. As
seen by the results in Fig. \ref{f:mhdrmn}, this complexity is clearly
evident.

Fig. \ref{f:mhdrmn} shows the final state of the fluid and magnetic
field at $t=80$. All waves shown in the results of Brio \& Wu are
evident, including the left moving fast rarefraction wave, slow shock
and rarefraction wave compound, right moving contact discontinuity,
slow shock and fast rarefraction wave. Comparisons also show that they
have moved with the correct velocities and magnitudes, and are in
close agreement with the published results of Brio \& Wu (\cite{BrioW88})
and Stone \& Norman (\cite{StoneN92b}).


\section{Summary}

In this paper we have described a numerical model for simulating
magnetised shear-flows in astrophysical systems such as galaxies or
accretions disks. Starting from the basic non-ideal MHD equations
describing the fluid, we have discussed their numerical implementation
on the standard shearing box model. Using explicit finite-difference
calculations, fluid properties are discretised onto a regular mesh in
three dimensions. The fluid is advanced through time using a third
order accurate predictor-corrector scheme, which has been shown to
compare favourably to other advancing schemes.

An important feature of the model comes from the diffusion methods,
which are implemented for two reasons, firstly to resolve and model
discontinuities in the flow, and secondly to stabilize the
numerics. The method works well as introducing diffusion only in the
necessary regions leaving well resolved regions of the fluid
unaffected.

An important aim of this work is to perform high resolution
calculations and as a result the model has been designed to take
advantage of parallel computers. We have therefore illustrated the
methods we have invoked to parallelise the code and shown their
performance. Similarly important feature of the model is to make it
easily adaptable to incorporate new physics or numerical
techniques. The parallelisation method adopted, namely HPF, has been
chosen for its flexibility and the ease of data parallelisation on a
finite-differences method. Certain features of the model, such as the
vertical and shearing boundaries, add complexity to the
parallelisation. However, having performed tests up to 128 processors,
the code has been shown to parallelise well, enabling high resolution
calculations within attainable times.

As well as giving the details of the methods we use, the other main
aim of this paper has been to justify the use of this code in future
simulations. We have performed several standard hydro- and
magnetohydrodynamic tests in a number of dimensions illustrating the
successes and limitation of the current method. We have shown that the
shock-capturing technique has performed well in a number of cases,
namely in modelling the Riemann shock tube problem and blast waves and
their interactions. In all of these test cases the shocks have
propagated at the correct speed and have shown profiles closely
matching the true ones. In the cases of spherical blast waves we have
also illustrated the capability of the method to produce spherical
structures on a Cartesian grid. The performed set of
magnetohydrodynamic tests have shown the capability of the code of
propagating Alfv\'en waves at correct velocities and shapes. These
tests have also shown the limitations of the use of magnetic vector
potential rather than the magnetic field itself such as in modelling
discontinuities in the field and maintaining the monotonicity of the
current. However, deviations from the physical solution are minor, and
considering the advantage of achieving solenoidal magnetic field, the
use of the model is justified.

As a first task we plan to use this method to investigate various
processes in the ISM, such as the nature of interstellar turbulence
driven by stellar explosive events, the structures emerging in such a
flow, leading to the investigation of dynamo processes in
galaxies. Another application is to study accretion processes in
accretion disks, in particular performing high-resolution calculations
to study in more detail the stresses acting within them. This model,
however, is far from complete, and additional physics can be added to
it. These include self-gravity for modelling molecular cloud formation
in the ISM, cooling functions and radiative transfer for accretion
disks. However in its present state a number of physical setups can be
investigated with this model, hopefully yielding interesting and
reliable information.

\begin{acknowledgements}
We wish to thank Prof. Ilkka Tuominen for his support and the valuable
comments during the work. We also acknowledge the Center for
Scientific Computing (CSC) in Espoo, Finland, for the parallelisation
assistance and computing time.
\end{acknowledgements}

\appendix
\section{Finite difference derivative operators}
\label{a:findif}

Centred, sixth order accurate, explicit finite difference operators
are used for all the derivatives (first and second) of the
pdes. Simple second order equivalents are used for the diffusive
quantities. All values are co-located on each gridpoints. The
derivatives take into account boundary conditions also which are also
based on a centred template.

The centred scheme is identical for all three dimensions and presented
below are the first and second derivatives of a variable $f$ located
at position $i,j,k$ taken in the $x$-direction. Indices are changed
accordingly for alternative directions.

The first derivative is calculated by:
\begin{eqnarray}
\frac{\partial f_{i,j,k}}{\partial x} = \frac{1}{\Delta x} 
[&\frac{3}{4}& (f_{i+1,j,k} - f_{i-1,j,k}) \nonumber \\
-&\frac{3}{20}& (f_{i+2,j,k} - f_{i-2,j,k}) \nonumber  \\
+&\frac{1}{60}& (f_{i+3,j,k} - f_{i-3,j,k})].
\end{eqnarray}

\noindent
The centred second derivative is as follows:
\begin{eqnarray}
\frac{\partial^2f_{i,j,k}}{\partial x^2} = \frac{1}{\Delta x^2} 
[-&\frac{49}{18}&f_{i,j,k} \nonumber \\
+&\frac{3}{2}& (f_{i+1,j,k} + f_{i-1,j,k})\nonumber  \\
-&\frac{3}{20}& (f_{i+2,j,k} + f_{i-2,j,k}) \nonumber \\
+&\frac{1}{90}& (f_{i+3,j,k} + f_{i-3,j,k})].
\end{eqnarray}

\section{time-stepping procedures}
\label{a:time-steps}

The Adams-Bashforth-Moulton predictor-corrector scheme consists of a
second order Adams-Bashforth predictor stage followed by the third
order Adams-Moulton corrector. Both are adapted for use with a
variable time-step.

The predictor stage is used to calculate the second order accurate
updated variable, $f_{n+1}^*$, at time $t_{n+1}=t_n+\Delta
t_{n}$. From this predicted value a derivative of the function may be
obtained at the new time and used again to calculate the third-order
corrected value $f_{n+1}$.

The second-order Adams-Bashforth predictor uses the current function
value along with its calculated derivative and the derivative from the
previous time-step and is defined as
\begin{equation}
f_{n+1}^*=f_n+\Delta t_n(af'_n+bf'_{n-1}),
\end{equation}

\noindent where prime denotes derivative with respect to time and $a$
and $b$ are constant coefficients to be determined. The key to
obtaining the second order accurate predicted value is to use Taylor
expansions around $t=t_n$ for $f$ and $f'$ from which $a$ and $b$ can
be determined such that only terms of the order of $\Delta t_n^3$
remain.

After substitution of $f_{n+1}$ and $f'_{n-1}$ in expanded form,
\begin{eqnarray}
f_{n+1} &=& f_n + \Delta t_n f'_n + \frac{\Delta t_n^2}{2}f''_n
+ \frac{\Delta t_n^3}{6}f'''_n+\cdots, \nonumber \\
f'_{n-1} &=& f'_n - \Delta t_{n-1} f''_n + \frac{\Delta t_{n-1}^2}{2}f'''_n
- \frac{\Delta t_{n-1}^3}{6}f''''_n+\cdots, \nonumber 
\end{eqnarray}

\noindent some straight forward linear algebra to remove terms up to the order
of $\Delta t_n^2$ yields
\begin{eqnarray}
a&=&-\frac{1}{2}\frac{\Delta t_n}{\Delta t_{n-1}}, \nonumber \\
b&=&1+\frac{1}{2}\frac{\Delta t_n}{\Delta t_{n-1}}, \nonumber 
\end{eqnarray}

If we define $r=\Delta t_n/\Delta t_{n-1}$ then the full second-order
Adams-Bashforth predictor step for variable time-steps is given by
\begin{equation}
f_{n+1}^*=f_n
+\Delta t_n \left(1+\frac{r}{2}\right)f'_n -\Delta t_n \frac{r}{2}f'_{n-1}.
\end{equation}

From the predicted value of $f_{n+1}^*$ we can calculate the predicted
derivative with respect to time at $t=t_{n+1}$ denoted as $f'_{n+1}$
for convenience here. This can then be used again to calculate the
corrected value $f_{n+1}$. The Adams-Moulton third-order corrector
stage is then defined as
\begin{equation}
f_{n+1}=f_n+\Delta t_n(af'_{n+1}+bf'_n+cf'_{n-1}).
\end{equation}

Again using Taylor expansions, values of $a$, $b$ and $c$ can be
calculated such that only terms of the order of $\Delta t_n^4$
remain. Using exactly the same analysis as above one finds that
\begin{eqnarray}
a&=&\frac{1}{2}\left(1-\frac{1}{3(1+r)}\right), \nonumber \\
b&=&\frac{1}{2}\left(1+\frac{1}{3r}\right), \nonumber \\
c&=&-\frac{1}{6r(1+r)}, \nonumber
\end{eqnarray}

\noindent yielding the final third-order Adams-Moulton corrector to be
\begin{eqnarray}
f_{n+1}=f_n&+&\frac{\Delta t_n}{6}\left(\frac{2+3r}{1+r}\right)f'_{n+1}
+\frac{\Delta t_n}{6}\left(\frac{1+3r}{r}\right)f'_n \nonumber \\
&-&\frac{\Delta t_n}{6}\frac{1}{r(1+r)}f'_{n-1}.
\end{eqnarray}

Both the predictor and corrector steps require knowledge of the
derivative at the last time-step. Obviously this is unknown for the
initial step. For this reason simulations are started from a second
order version of the Adams-Bashforth-Moulton scheme (first order
predictor, second order corrector). These are calculated in exactly
the same way as above but terms including $f'_{n-1}$ are
neglected. Hence one obtains the predictor and corrector to be respectively
\begin{eqnarray}
f^*_{n+1} &=& f_n+\Delta t_n f'_n, \\
f_{n+1}&=& f_n+\frac{\Delta t_n}{2}(f'_{n+1}+f'_n).
\end{eqnarray}

After the first time-step, all subsequent steps may be performed with
the third order method using the derivative stored from the previous
time-step.



\begin{thebibliography}{}
\bibitem[1991]{BH91}
  Balbus, S. A., Hawley, J. F. 1991, ApJ, 376, 214

\bibitem[1999]{Balsara99}
  Balsara D., Pouquet A. 1999, Phys. Plasmas, 6, 89

\bibitem[2000]{Aksu2000}
  Brandenburg A. 2000, in press, Astro-ph/0006186

\bibitem[1995]{BNST95}
  Brandenburg, A., Nordlund, \AA., Stein, R.~F., Torkelsson, U., \
  1995, ApJ, 446, 741

\bibitem[1988]{BrioW88}
  Brio, M., Wu, C.~C. \ 1988, J. Comput. Phys., 75, 400

\bibitem[1996]{Hughes96}
  Cattaneo, F., Hughes, D. W. 1996, Phys. Review E, 54, 4532

\bibitem[1948]{CourantF48}
  Courant, R., Friedrichs, K.~O. \ 1948, {\em Supersonic Flow and
Shock Waves}, New York

\bibitem[1972]{dalgarnomc72}
  Dalgarno, A., McCray, R.~A.\ 1972, ARA\&A, 10, 375

\bibitem[1991]{FerriereMMZ91}
  Ferri\'ere, K. M., Mac Low, M.-M., Zweibel, E. G. 1991, ApJ, 375, 239

\bibitem[1975]{Frisch75}
  Frisch, U., Pouquet, A., L\'eorat, J., Mazure, A. 1975, JFM, 68, 769  

\bibitem[1996]{GalsgaardN96}
  Galsgaard, K., Nordlund, \AA. \ 1996, JGR, 101, 445

\bibitem[2000]{HawleyK00}
  Hawley, J. F., Krolik, J. H. 2000, in press, Astro-ph/0006456

\bibitem[1984]{HawleySW84}
  Hawley, J. F., Smarr, L. L., Wilson, J. R. 1984, ApJ, 277, 296 

\bibitem[1995]{HawleyGB95}
  Hawley, J. F., Gammie, C. F., Balbus, S. A. 1995, ApJ, 440, 742

\bibitem[1987]{Heiles87}
  Heiles, C. 1987, ApJ, 315, 555

\bibitem[1979]{Hyman79}
  Hyman, J.~M.\ 1979, in Adv. in Computer Methods for Partial
  Differential Equations, Vol. III, ed R. Vichnevetsky \&
  R.S. Stepleman (Publ. IMACS), 313

\bibitem[1966]{JT66}
  Julian, W. H., Toomre, A. 1966, ApJ, 146, 810

\bibitem[1992]{Koo92}
  Koo, B.-C., Heiles, C., Reach, W. T. 1992, ApJ, 390, 108

\bibitem[1999]{Korpi99}
  Korpi, M. J., Brandenburg, A., Shukurov, A., Tuominen, I. 1999, ApJ,
  514, L99
  
\bibitem[1999]{MacLow99}
  Mac Low, M.-M. 1999, ApJ, 524, 169

\bibitem[1980]{MousP80}
  Mouschovias, T.~Ch., Paleologou, E.~V. \ 1980, ApJ, 237, 877

\bibitem[1950]{vonNeumannR50}
  von Neumann, J., Richtmyer, R.~D. \ 1950, J. Appl. Phys., 21, 232

\bibitem[1997]{NordlundG97}
  Nordlund, \AA., Galsgaard, K.\ 1997, Tech. Rep., Astronomical
  Observatory, Copenhagen University, 1997

\bibitem[1990]{NordlundS90}
  Nordlund, \AA., Stein, R.~F.\ 1990, Comput.\ Phys.\ Comm., 59, 119

\bibitem[1996]{Norma96}
  Normandeau, M., Taylor, A. R., Dewdney, P. E. 1996, Nature, 380, 687

\bibitem[1988]{OM88}
  Ostriker, J. P., McKee, C. F. 1988, Rev. Mod. Phys., 60, 1

\bibitem[1955]{Parker55}
  Parker, E. N. 1955, ApJ, 122, 293

\bibitem[1976]{inverse76}
  Pouquet, A., Frisch, U., L\'eorat, J. 1976, JFM, 77, 321 

\bibitem[1976]{RaymondCS76}
  Raymond, J.~C., Cox, D.~P., Smith, B.~W.\ 1976, ApJ, 204, 290

\bibitem[1995]{RosenB95}
  Rosen, A., Bregman, J.~N.\ 1995, ApJ, 440, 634

\bibitem[1959]{Sedov59}
  Sedov, L. I. 1959, Similarity and dimensional methods in mechanics,
  Infosearch Ltd., London

\bibitem[1992]{Shu92}
  Shu, F. H. 1992, The physics of astrophysics, Volume II: Gas
  dynamics, University Science Books, California 

\bibitem[1953]{SS53}
  Spitzer, L. Jr., Schwarzschild, M. 1953, ApJ, 118, 106

\bibitem[1978]{Sod78}
  Sod, G.~A. \ 1978, J. Comput. Phys., 27, 1

\bibitem[1966]{SKR66}
  Steenbeck, M., Krause, F., R\"adler, K.-H. 1966, Z. Naturforsch. 21a,
  369 

\bibitem[1998]{SteinN98}
  Stein, R.~F., Nordlund, \AA \
  1998, ApJ, 499, 914

\bibitem[1992a]{StoneN92}
  Stone, J.~M., Norman, M.~L. \ 1992a, ApJS, 80, 753

\bibitem[1992b]{StoneN92b}
  Stone, J.~M., Norman, M.~L. \ 1992b, ApJS, 80, 791

\bibitem[1992]{StoneHEN92}
  Stone, J.~M., Hawley, J.~F., Evans, C.~R., Norman, M.~L. \ 
  1992, ApJ, 388, 415

\bibitem[1998]{Tomisaka98}
  Tomisaka K. 1998, MNRAS, 298, 797

\bibitem[1950]{Taylor50}
  Taylor, G. I. 1950, Proc. Roy. Soc., Ser. A, 201, 175

\bibitem[1995]{VazquezP95}
  V\'azquez-Semadeni, E., Passot, T., Pouquet, A.\ 1995, ApJ,
  441, 702

\bibitem[1995]{VoinovichC95}
  Voinovich, P.~A., Chernin, A.~D. \ 1995 Astronomy Letters, 21, 835

\bibitem[1988]{WisdomT88}
  Wisdom, J., Tremaine, S. 1988, AJ, 95, 925

\bibitem[1984]{WoodwardC84}
  Woodward, P., Colella, P. \ 1984, J. Comput. Phys., 54, 115

\bibitem[1990]{YoshiokaI90}  
  Yoshioka, S., Ikeuchi S. 1990, ApJ, 360, 352

\end{thebibliography}
\end{document}